\documentclass[aps,showpacs]{revtex4}
\usepackage{epsfig}
\usepackage{graphicx}
\begin{document}
\title{Modulated wavepackets
associated with longitudinal dust grain oscillations \\ in a dusty
plasma crystal
 \footnote{Preprint, submitted to \it{Physics of Plasmas}.}}
\author{I. Kourakis}
\altaffiliation[On leave from: ]{U.L.B. - Universit\'e Libre de
Bruxelles, Facult\'e des Sciences Apliqu\'ees - C.P. 165/81
Physique G\'en\'erale, Avenue F. D. Roosevelt 49, B-1050 Brussels,
Belgium; also at: U.L.B., Association Euratom -- Etat Belge,
C.P. 231
Physique Statistique et Plasmas, Boulevard du Triomphe, B-1050 Brussels,
Belgium.}

\email{ioannis@tp4.rub.de}
\author{P. K. Shukla}
\email{ps@tp4.rub.de} \affiliation{Institut f\"ur Theoretische
Physik IV, Fakult\"at f\"ur Physik und Astronomie,
Ruhr--Universit\"at Bochum, D-44780 Bochum, Germany}
\date{Received 5 December 2003; accepted 13 January 2004}

\begin{abstract}
The nonlinear amplitude modulation of longitudinal dust lattice
waves (LDLWs) propagating in a dusty plasma crystal is
investigated in a continuum approximation. It is shown that long
wavelength LDLWs are modulationally stable, while shorter
wavelengths may be unstable. The possibility for the formation and
propagation of different envelope localized excitations is
discussed. It is shown that the total grain displacement bears a
(weak) constant displacement (zeroth harmonic mode), due to the
asymmetric form of the nonlinear interaction potential. The
existence of \emph{asymmetric} envelope localized modes is
predicted. The types and characteristics of these coherent
nonlinear structures are discussed.
\end{abstract}

\maketitle

\section{Introduction}

A remarkable new feature associated with the physics of
dust-contaminated plasmas \cite{PSbook, Verheest} is the existence
of strongly-coupled charged matter configurations, which even
lead to the formation of dusty plasma (DP) Wigner-type layers
(crystals) when the inter-grain potential energy far exceeds the
average kinetic energy of the dust particles. Dust crystals display
a variety of new phenomena, such as phase transitions
(crystallization, melting) and lattice excitations;  a link is
thus established between plasma physics and solid state physics
\cite{Kittel}.

Dust lattices, which are typically formed in the sheath region in
discharge experiments and remain suspended above the negative
electrode due to a balance between the external electric and
gravity forces \cite{Chu}, are known to support harmonic
oscillations (acoustic modes) in both longitudinal and
transverse-shear (horizontal-plane) directions, as well as
optical-mode-like oscillations in the vertical (off-plane)
direction \cite{Melandso, farokhi, tskhakaya, Wang, Morfill,
Nunomura2}.

Dust-lattice waves (DLWs) are reminiscent of waves (`phonons')
propagating in atomic chains, which are long known to be dominated
by interesting nonlinear phenomena (localized modes,
instabilities), due to the intrinsic nonlinearities of
inter-atomic interaction mechanisms and/or on-site substrate
potentials \cite{Tsurui, Flytzanis1985, Peyrard, Scott}. The
present study is devoted to the study of one such phenomenon,
namely the nonlinear amplitude modulation of weakly nonlinear
oscillations with respect to longitudinal lattice oscillations.
Even though this mechanism, which is associated with harmonic
generation due to self-interaction of the carrier waves, is today
known to be of relevance in phenomena as diverse as the modulational
instability and energy localization in solids, charge and
information transport in biomolecules and DNA strands, and coherent
signal transmission in electric lines and in optical pulse
propagation \cite{Davydov, Remoissenet, Newell2, Scott, Peyrard,
Hasegawa1}, it has only recently been considered in dusty plasmas
\cite{IKPKSPoP} with regard to the propagation of dust-acoustic
and dust-ion acoustic waves \cite{PSbook, Verheest} in
a weakly-coupled (gas-like) DP. As far as the nonlinear modulation
of dust-lattice oscillations is concerned, our knowledge is still
at a very early stage, let alone some first attempts to model the
modulation of longitudinal \cite{AMS2} and transverse
\cite{IKPKSPhScr} DLWs. In addition to a theoretical investigation
of the occurrence of modulational instability, these studies have
provided a first prediction of the possible existence of localized
envelope excitations in a dust-crystal. It should be stressed
that, as long known and extensively studied in solid state
physics, the physics of these localized modes (basically associated with
the generation of harmonics, which was recently observed experimentally;
see in Ref. \cite{Nunomura}) is essentially
different from that of compressive solitons, described by
Korteweg-deVries (KdV)-- type models, first introduced by
Melands\o  \ \cite{Melandso} and later considered in some experiments
\cite{Nosenko, Samsonov} and in theoretical papers
\cite{Samsonov, Avinash, Zhdanov, PKS2003}.

In this paper, we consider the amplitude modulation of
longitudinal waves propagating in DP lattices. We will derive an
equation governing the evolution of the dust lattice wave envelope
and will then examine its stability with respect to external
perturbations. Explicit conditions for the formation of different
types of envelope excitations will be presented, in terms of the
physical parameters involved in DP crystals and, in particular,
the existence of \emph{asymmetric} envelope dust-lattice solitons
is predicted. This study refers to the longitudinal grain motion
in an `ideal' one--dimensional DP crystal, i.e. a single,
unidimensional, infinite-sized, dust-layer of identical (in size,
charge and mass) dust grains situated at spatially periodic sites
(at equilibrium). It should however be noted that real crystals
formed in the laboratory are often two-- or three--dimensional
arrays \cite{Morfill, Kong}, which may host longitudinal waves
such as the ones considered herein, but also transverse (shear)
waves \cite{Nunomura2} as well as dissipative nonlinear
structures, as observed in recent experiments \cite{Nunomura,
Samsonov}. Two--dimensional (2D) lattices are known from Solid
State Physics to possess an enriched profile of possible
excitations, including e.g. oblique (with respect to the principal
axes) waves, 2D breathers and vortex excitations \cite{Tamga}.
Therefore, this work is only a first step towards the
investigation of nonlinear mechanisms in DP crystals: the
mechanism put forward in this paper may be generalized to higher
dimensionality; nevertheless, this is left for consideration in
future work.

\section{The model}

Let us consider a layer of charged dust grains (mass $M$ and charge
$q$, both assumed constant for simplicity) of lattice constant
$r_0$. The Hamiltonian of such a chain is of the form
\[H = \sum_n \frac{1}{2} \, M \, \biggl( \frac{d \mathbf{r}_n}{dt} \biggr)^2 \, + \,
\sum_{m \ne n} U(r_{nm}) \, ,
\]
where $\mathbf{r}_n$ is the position vector of the $n-$th grain;
$U_{nm}(r_{nm}) \equiv q \, \phi(x)$ is a binary interaction
potential function related to the electrostatic potential
$\phi(x)$ around the $m-$th grain, and $r_{nm} =
|\mathbf{r}_{n}-\mathbf{r}_{m}|$ is the distance between the
$n-$th and $m-$th grains. We shall limit ourselves to considering
the {\em longitudinal} ($\sim \hat x $) motion of the $n-$th dust
grain, which obeys
\begin{equation}
M\, \biggl( \frac{d^2 x_n}{dt^2} \, + \nu \, \frac{d x_n}{dt}
\biggr) = \,- \sum_n \,\frac{\partial U_{nm}(r_{nm})}{\partial
x_n} \, \equiv \, q \, E(x_n), \label{eqmotion0}
\end{equation}
where $E(x) = - \partial \phi(x)/\partial x$ is the electric
field; the usual \textit{ad hoc} damping term is introduced in the
left-hand-side of Eq. (\ref{eqmotion0}), involving the damping
rate $\nu$ due to dust--neutral collisions. A
one-dimensional (1D) DP layer is considered here, but the
generalization to a two-dimensional (2D) grid is straightforward.
At a first step, we have omitted the external force term
$F_{ext}$, often introduced to account for the initial laser
excitation and/or the parabolic confinement which ensures
horizontal lattice equilibrium in experiments \cite{Samsonov}.

\subsection{Equation of motion}

Assuming small displacements from equilibrium, one may Taylor
expand the interaction potential $\phi(r)$ around the equilibrium
inter-grain distance $l r_0 =  |n-m| r_0$ (between $l-$th order
neighbors, $l=1, 2, ...$), viz.
\[\phi(r_{nm}) = \sum_{l'=0}^\infty \,\frac{1}{l'!}
\biggl. \frac{d^{l'} \phi(r)}{d r^{l'}} \biggr|_{r = |n-m| r_0} \,
(x_{n} - x_{m})^{l'} \, ,\] where $l'$ denotes the degree of
nonlinearity involved in its contribution: $l' = 1$ is the linear
interaction term, $l' = 2$ stands for the quadratic potential
nonlinearity, and so forth. Obviously, $\delta x_n =  x_n -
x_n^{(0)}$ denotes the displacement of the $n-$th grain from
equilibrium, which now follows
\begin{eqnarray}
M\, \biggl[ \frac{d^2 (\delta x_n)}{dt^2} \, + \nu \, \frac{d
(\delta x_n)}{dt} \biggr] &=& \, q \biggl\{ \phi''(r_0) \, (\delta
x_{n+1} + \delta x_{n-1} - 2 \delta x_{n}) \nonumber \\
& & \, + \sum_{l'=2}^\infty \frac{1}{l'!}\,\biggl. \frac{d
\phi^{l'+1}(r)}{d r^{l'+1}} \biggr|_{r = r_0}  \, \bigl[ (\delta
x_{n+1} - \delta x_{n})^{l'} -  (\delta x_{n} - \delta
x_{n-1})^{l'} \bigr]
 \nonumber \\
 & & \, + \sum_{l=2}^N \,\phi''(l r_0) \, (\delta x_{n+l} + \delta
x_{n-l} - 2 \delta x_{n})
\nonumber \\
& & \, + \sum_{l = 2}^N \sum_{l' = 2}^\infty
\frac{1}{l'!}\,\biggl. \frac{d \phi^{l'+1}(r)}{d r^{l'+1}}
\biggr|_{r = l r_0}  \, \biggl[ (\delta x_{n+l} - \delta
x_{n})^{l'} -  (\delta x_{n} -
\delta x_{n-l})^{l'} \biggr] \biggr\} \, . \nonumber \\
 \label{eqmotion1}
\end{eqnarray}
We have distinguished the linear/nonlinear contributions of the
first neighbors (1st/2nd lines) from the corresponding longer
neighbor terms (3rd/4th lines, respectively).

Adopting the standard continuum approximation, often employed in
solid state physics \cite{Kittel}, we may assume that only small
displacement variations occur between neighboring sites, i.e.
\[
\delta x_{n \pm l} = \delta x_{n} \pm l r_0 \frac{\partial
u}{\partial x} + \frac{1}{2} (l r_0)^2 \frac{\partial^2
u}{\partial x^2} \pm \frac{1}{3!} (l r_0)^3 \frac{\partial^3
u}{\partial x^3} + \frac{1}{4!} (l r_0)^4 \frac{\partial^4
u}{\partial x^4} \pm \,  ... ,
\]
where the displacement $\delta x (t)$ is now expressed by a
continuous function $u = u(x, t)$. One may now proceed by
inserting this ansatz in the discrete equation of motion
(\ref{eqmotion1}), and carefully evaluating the contribution of
each term. The calculation, quite tedious yet perfectly
straightforward (to be reported, in full detail, elsewhere
\cite{IKPKS}), leads to a continuum equation of motion in
the form
\begin{equation}
\ddot{u}  \,+ \, \nu \, \dot{u} - v_0^2 \, u_{xx}\,= \, v_1^2\,
r_0^2 \, u_{xxxx}\, - \, p_0 \, u_x \,u_{xx} \, + \, q_0 \, (u_x)^2 \,u_{xx} \, ,
\label{eqmotion-gen-continuum}
\end{equation}
where the subscript $x$ denotes the partial differentiation, viz. $u_x
\, u_{xx} = (u_x^2)_x/2$ and $(u_x)^2 \, u_{xx} = (u_x^3)_x/3$.

\subsection{Characteristic physical quantities}

The definition of the `sound speed', $v_0 = \omega_{0, L}
\,r_0$, associated with the longitudinal oscillation
eigenfrequency $\omega_{0, L}$, is
\begin{equation} v_0 = \frac{q}{M} \,r_0^{2}
\sum_{l = 1}^N \phi''(l r_0)\, l^2 \,\equiv \, \omega_{0, L}^2
\,r_0^2 \, ,\label{defv0}
\end{equation}
where $N$ is the degree of interacting site vicinity assumed:
$N=1$ for first-neighbor interactions (FNI), $N=2$ for
second-neighbor interactions (SNI), and so forth. The dispersion
coefficient, involving the characteristic velocity $v_1$, is given
by
\begin{equation} v_1^2 \,
r_0^2  = \frac{1}{12} \frac{q}{M} \,r_0^{4} \sum_{l = 1}^N
\phi''(l r_0)\, l^4 \,. \label{defv1}
\end{equation}
We note that $v_1^2 = v_0^2/12$ for $N=1$ (only). The quadratic
force (or cubic potential) nonlinearity coefficient is
\begin{equation}  p_0 = - \frac{q}{M} \,r_0^{3}
\sum_{l = 1}^N \phi'''(l r_0)\, l^3 \, . \label{defc11}
\end{equation}
Finally, the cubic force (or quartic potential) nonlinearity
coefficient $q_0$ is
\begin{equation} q_0 = \frac{1}{2} \frac{q}{M} \,r_0^{4}
\, \sum_{l = 1}^N \phi''''(l r_0)\, l^4 \, . \label{defc111}
\end{equation}

\begin{figure}[htb]
 \centering
 \resizebox{4in}{!}{
 \includegraphics[]{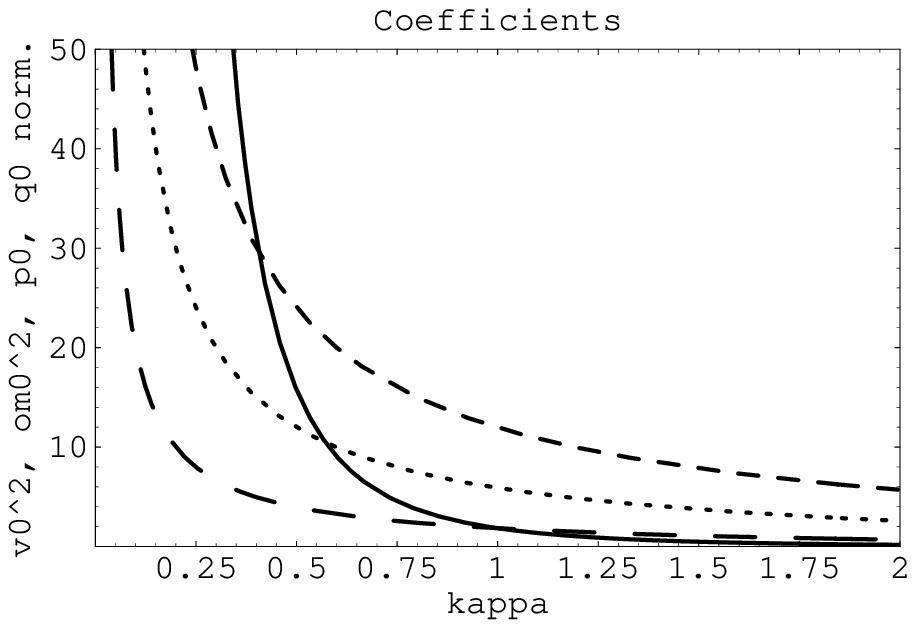}} \\
\vskip 2 cm \resizebox{4in}{!}{
 \includegraphics[]{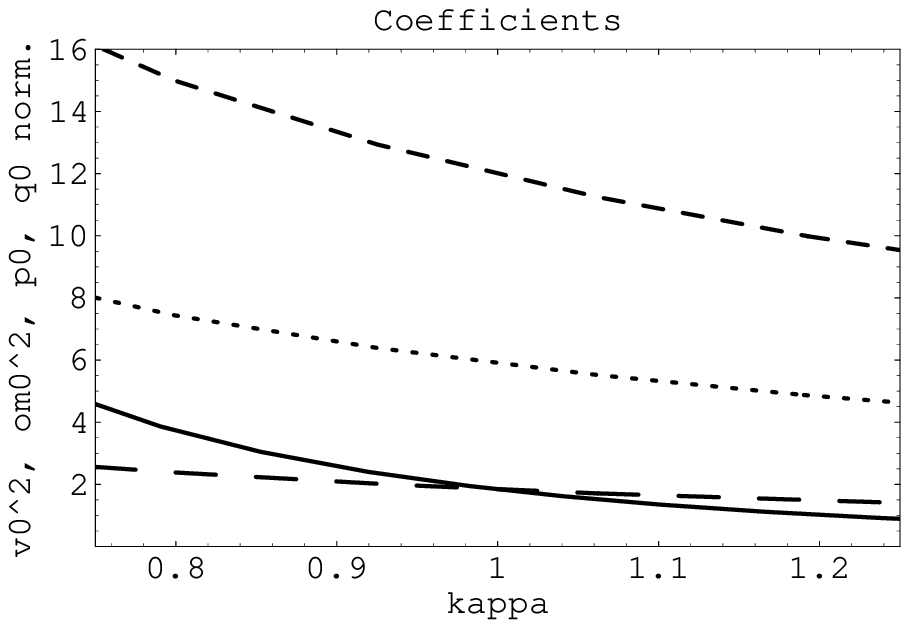}}
\caption{} \label{figure1}
\end{figure}

Let us point out that the above  definitions of the coefficients
in (\ref{eqmotion-gen-continuum}) are inspired by the
Debye--H\"uckel (Yukawa) potential of the form $\phi_{D}(r) = (q/r)\,
e^{-r/\lambda_D}$ (whose odd/even derivatives are
negative/positive), in which case they are defined in such a way
that all of $v_0^2$, $v_1^2$, $p_0$ and $q_0$ take {\em{positive}}
values. In this case (and for $N=1$ i.e. FNI), relations
(\ref{defv0}), (\ref{defv1}), (\ref{defc11}) and (\ref{defc111})
yield
\begin{equation}
{\omega_{L, 0}^2} = \frac{2 q^2}{M \lambda_D^3} \, e^{-\kappa}\,
\frac{1 + \kappa + \kappa^2/2}{\kappa^3}\, = \,
\frac{v_0^2}{\kappa^2 \lambda_D^2} = 12 \frac{v_1^2}{\kappa^2
\lambda_D^2} \, , \label{Debye-om0}
\end{equation}
\begin{equation}
p_0 = \frac{6 q^2}{M \lambda_D} \, e^{-\kappa}\, \biggl(
\frac{1}{\kappa} +  1 + \frac{\kappa}{2}+ \frac{\kappa^2}{6}
\biggr)\, , \label{Debye-p0}
\end{equation}
\begin{equation}
q_0 = \frac{12 q^2}{M \lambda_D} \, e^{-\kappa}\, \biggl(
\frac{1}{\kappa} +  1 + \frac{\kappa}{2}+ \frac{\kappa^2}{6} +
\frac{\kappa^3}{24} \biggr)\, , \label{Debye-q0}
\end{equation}
where we have introduced the \textit{lattice parameter} $\kappa =
r_0/\lambda_D$; in fact, $\kappa$ is roughly between $1$ and $1.5$
in DP chains spontaneously formed in current dusty plasma
discharges \cite{Chu, Nosenko, Samsonov}. These relations, which
are depicted in Fig. \ref{figure1}, coincide with the ones in
previous studies for FNI \cite{Melandso, AMS2}. Specifically,
compared to the notation of Melands\o  \
[see (15), (16), (26) in
Ref. \cite{Melandso}], $v_0^2$, $v_1^2$ and $p_0$ above correspond
to $v_0^2$, $v_0^2/12$ and $\gamma(a) a^3/M$ therein. Also,
compared to the notation of Amin \textit{et al.} [see (1)--(4) in
Ref. \cite{AMS2}], $v_0^2$, $v_1^2 r_0^2$ and $p_0$ above
correspond to $\alpha_0$, $\beta_0$ and $2 \gamma_0$ therein.
However, the coefficient $q_0$, which can easily be seen to be
practically $\approx 2 p_0$ or higher (see Fig. \ref{figure1}) --
i.e. too important to be omitted -- is neglected in those studies.

One should retain, for later reference,
the form of the discrete dispersion relation, obtained from the general
(discrete) equation of motion (\ref{eqmotion1})
\begin{equation}
\omega \, ( \omega + i \, \nu ) \, = \, \frac{4 q}{M}\, \sum_{l =
1}^N\, \phi''(l r_0)\, \sin^2\biggl( \frac{l \, k r_0}{2}\biggr)
\, =  \, \frac{4 q}{M}\, \sum_{l = 1}^N\, \phi''(l \kappa
\lambda_D)\, \sin^2\biggl[\frac{l \, \kappa \,(k
\lambda_D)}{2}\biggr] \, . \label{dispersion-discrete}
\end{equation}
One may readily verify that the standard 1D acoustic wave
dispersion relation $\omega \approx k\,v_0$ is obtained in the
small $k$ (long wavelength) limit: check by setting $\sin({l \, k
r_0}/{2}) \approx {l k r_0}/{2}$ (and recalling the general
definition of $v_0$ above). For FNI and Debye--H\"uckel
interactions, one obtains
\begin{equation}
\omega \, ( \omega + i \, \nu ) \, = \, \frac{4 q}{M}\,
\phi''(r_0)\, \sin^2\frac{k r_0}{2} \, \equiv \,4 \, \omega_{L, 0}^2 \,
\sin^2\frac{k r_0}{2} \,  \label{dispersion-approx}
\end{equation}
[cf. (\ref{defv0}) for $N = 1$].
We note that the LDLW frequency and the sound speed will be of
the order of: $\hat \omega_0 = [2 q^2/(M \lambda_D^3)]^{1/2}$
($\equiv \omega_0 \sqrt{2} = \omega_{max}$) and $\hat v_0 = [2
q^2/(M \lambda_D)]^{1/2} \equiv \omega_0 \lambda_D^2$,
respectively; these characteristic quantities scale as $\hat
\omega_0 \sim \,r^{-1/2}\,a^{-3/2}\,U$ and $\hat \omega_0 \sim
\,r^{-1/2}\,a^{-1/2}\,U$, respectively, for a given dust grain
radius $r$, an inter-grain separation $a$ and the dust surface
potential $U = q/r$ (following Ref. \cite{Melandso}).

Even though all the formulae above are provided in their general
form, in the following we shall assume that $N=1$ (FNI) and will
also set $\nu \approx 0$ at a first step, for the sake of
analytical tractability. These assumptions will be relaxed in a
forthcoming work, also accounting for higher-order nonlinearities
neglected here, which may be evaluated in a systematic manner
\cite{IKPKS}. Our aim here is to emphasize on a physical
mechanism, rather than putting forward a tedious and exhaustive
mathematical model.

\section{Multiple scale expansion - derivation of a Nonlinear Schr\"odinger
Equation}

According to the standard reductive perturbation method
\cite{redpert, Tsurui}, we shall consider a small displacement of
the form: \( u \rightarrow 0 + \epsilon \, u_1 + \epsilon^2 u_2 +
... \, , \) where $\epsilon \ll 1$ is a small parameter, and
solutions $u_n$ at each order are assumed to be a sum of $m-$th
order harmonics, viz. $u_n = \sum_{m=0}^n u_m^{(n)} \, \exp[i (k x
- \omega t)]$ (the reality condition $u_{-m}^{(n)} =
{u_m^{(n)}}^*$ is understood). Time and space scales are
accordingly expanded as
\[
\partial/\partial t \rightarrow \partial/\partial T_0 + \epsilon
\,
\partial/\partial T_1 + \epsilon^2 \partial/\partial T_2 + ...
\equiv \, \partial_0 + \epsilon \,
\partial_1 + \epsilon^2 \partial_2 + ... \, ,
\]
and
\[
\partial/\partial x \rightarrow \partial/\partial X_0 + \epsilon
\,
\partial/\partial X_1 + \epsilon^2 \partial/\partial X_2 + ...
\equiv \, \nabla_0 + \epsilon \, \nabla_1 + \epsilon^2 \nabla_2 +
... \, ,
\]
implying that
\[
\partial^2/\partial t^2 \rightarrow \, \partial_0^2 + 2 \, \epsilon \,
\partial_0 \partial_1 + \epsilon^2 (\partial_1^2 + 2 \partial_0 \partial_2) + ...
\, ,\]
\[
\partial^2/\partial x^2 \rightarrow \, \nabla_0^2 + 2\,  \epsilon \,
\nabla_0 \nabla_1 + \epsilon^2 (\nabla_1^2 + 2 \nabla_0 \nabla_2)
+ ... \, ,
\]
and
\[
\partial^4/\partial x^4 \rightarrow \, \nabla_0^4 + 4\,  \epsilon \,
\nabla_0^3 \nabla_1 + 2 \, \epsilon^2 \nabla_0^2\, (3 \nabla_1^2 +
2 \nabla_0 \nabla_2) + ... \, .
\]

We now proceed by substituting all the above series in
(\ref{eqmotion-gen-continuum}) and isolating terms arising in the
equation of motion at each order in $\epsilon^n$. By solving for
$u_n$, then substituting in the following order $\epsilon^{n+1}$
and so forth, we obtain the $m-$th harmonic amplitudes
$u_n^{(m)}$ in each order,
along with a compatibility condition up to any given order.

The equation obtained in order $\sim \epsilon^1$ is
\begin{equation}
(\partial_0^2 - v_0^2 \,\nabla_0^2 - v_1^2 r_0^2 \, \,\nabla_0^4)
\, u_1 \, \equiv \, L_0\, u_1 \,= 0 \, . \label{eq-epsilon1}
\end{equation}
We may assume that $u_1 = u_1^{(1)} \, \exp[i (k x - \omega t)] +
c.c. \equiv \exp (i \theta)  + c.c.$, where $\omega$, $k = 2
\pi/\lambda$ and $\lambda$ denote the carrier wave frequency,
wavenumber and wavelength, respectively; $c.c.$ stands for the
complex conjugate of the preceding quantity everywhere.

\begin{figure}[htb]
 \centering
 \resizebox{4in}{!}{
 \includegraphics[]{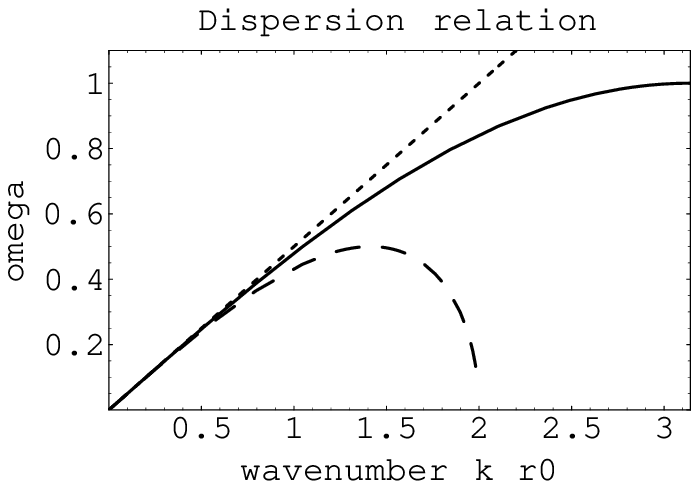}} \\
\vskip 2 cm \resizebox{4in}{!}{
 \includegraphics[]{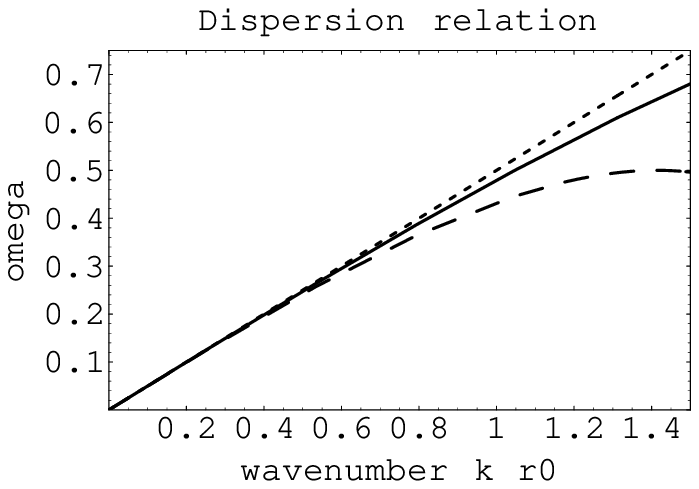}}
\caption{} \label{figure2}
\end{figure}

The dispersion relation obtained from (\ref{eq-epsilon1}) is of
the form
\begin{equation}
\omega^2 = k^2 \,v_0^2 \,  - k^4 \, v_1^2 \, r_0^2 \,  = k^2 \,v_0^2 \,
\biggl(1 - \, \frac{v_1^2}{v_0^2} k^2 \, r_0^2 \biggr) \, ,
\label{dispersion}
\end{equation}
predicting an acoustic behaviour (viz. $\omega \approx k v_0$) for
\emph{very} low $k$, and dispersion (weighted by $v_1^2 r_0^2$) at
higher values. Notice that the reality condition $\omega > 0$,
implying $k < (v_0/v_1) \, r_0^{-1} \equiv k_{cr, 1}$ ($=
\sqrt{12} \, r_0^{-1}$ for FNI), is in principle satisfied,
covered by the continuum hypothesis $k \ll (\pi/r_0) \equiv k_{cr,
0}$. Specifically, one may numerically verify the agreement
between the discrete dispersion relation in
(\ref{dispersion-approx}) (for $\nu = 0$) and its continuum
approximation (\ref{dispersion}) -- see Fig. \ref{figure2}; for
instance, aiming at an accuracy interval of 99 \% (90 \%), i.e.
allowing for a relative error of 1 \% (10 \%), one should restrict
the confidence interval of this study to values of $k r_0$ below
$0.3$ ($1.0$).
%In consequence, we will not adress the issues of negative $\omega^2$ or $v_g$,
%anymore.

Let us evaluate the action of the linear operator $L_0$, defined
above, on higher harmonics of the phase $\theta$
\begin{equation}
L_0\, e^{i n \theta} = [(-i n \omega)^2 - v_0^2 (i n k)^2 - v_1^2
\, r_0^2 (i n k)^4 ] \, e^{i n \theta} = ... = - n^2 (n^2 - 1)
v_1^2 \, r_0^2 \, k^4 \,e^{i n \theta} \equiv D_n \, e^{i n
\theta}\, , \label{Dn}
\end{equation}
where we made use of the dispersion relation (\ref{dispersion}).
We observe that $D_0 = D_1 = 0$.

To order $\sim \epsilon^2$, we have
\begin{equation}
L_0\, u_2 \,= - 2 (\partial_0 \,\partial_1 \, -v_0^2 \nabla_0
\nabla_1 \, - 2 v_1^2 \, r_0^2 \, \nabla_0^3 \nabla_1) \, u_1 \, -
p_0 \,(\nabla_0 u_1) \,(\nabla_0^2 u_1)\, . \label{eq-epsilon2}
\end{equation}
One now has to impose the condition of suppression of secular terms
(i.e. terms $\sim e^{i \theta}$ in the right--hand side, which may
resonate with the null space of the operator $L_0$), which takes
the form
\begin{equation}
\frac{\partial u_1^{(1)}}{\partial T_1 } + v_g\, \frac{\partial
u_1^{(1)}}{\partial X_1 } = 0\, ,  \label{secular1}
\end{equation}
or ${\partial u_1^{(1)}}/{\partial \zeta} = 0$, i.e. $u_1^{(1)} =
u_1^{(1)}(\xi, \tau)$, where $\zeta = \epsilon (x - v_g t)$ and
$\tau = \epsilon^2 t$, implying that the slowly--varying
wave--front (envelope) travels at the \textit{group velocity} $v_g
= \omega'(k) =(v_0^2 - 2 v_1^2 r_0^2 k^2) k/\omega$. We notice
that $v_g$ becomes zero at $k = v_0/(v_1 \sqrt{2})\, r_0^{-1}
\equiv k_{cr, 1}/\sqrt{2}$ ($= \sqrt{6} \,r_0^{-1}$ for FNI),
beyond which it becomes negative: the wave envelope and the
carrier wave then propagate in opposite directions (backward
wave). Nevertheless, for the sake of rigor, note that this comment
rather becomes obsolete once one properly takes into account the
influence of the lattice discreteness on the form of the
dispersion relation $\omega = \omega(k)$; see Fig. \ref{figure2}.
%, shifting the position of the maximum, viz. $\omega' = v_g = 0$,
% towards higher $k$.

The solution $u_2$ obtained at this order involves a second
harmonic contribution $u_2^{(2)}\,\exp(2i\theta) = (i p_0 k^3/D_2)
({u_1^{(1)}})^2 \,\exp(2i\theta) = - i [p_0/(12 v_1^2 r_0^2 k)]
\,({u_1^{(1)}})^2 \exp(2i\theta)$, as well as a zeroth (and
possibly a first) harmonic(s), which may be determined by
higher-order relations.

One may now proceed to the next (third) order in $\epsilon$. The
condition for annihilation of secular terms now takes the form
\begin{equation}
i \frac{\partial u_1^{(1)}}{\partial \tau} \, + \, P \,
\frac{\partial^2 u_1^{(1)}}{\partial \zeta^2} \, + \,Q_0 \,
|u_1^{(1)}|^2 u_1^{(1)}\, + \frac{p_0 k^2}{2 \omega} \, u_1^{(1)}
\frac{\partial u_0^{(1)}}{\partial \zeta} = 0\, . \label{NLSE01}
\end{equation}
The \textit{dispersion coefficient} $P$ is
\begin{equation}
P \, = \,\frac{1}{2} \omega''(k) = \,-\frac{1}{2 \omega} (v_g^2 -
v_0^2 + 6 v_1^2 r_0^2 k^2) = \frac{v_1^4 r_0^4 k^4}{\omega^3}\,
\biggl(k^2 - \frac{3 v_0^2}{2 v_1^2 r_0^2} \biggr) \, ,
\label{Pcoeff}
\end{equation}
while the \textit{nonlinearity coefficient} $Q_0$ is
\begin{equation}
Q_0 \, = \,- \frac{k^2}{2 \omega} \, \biggl(q_0 \,k^2 +
\frac{p_0^2}{6 v_1^2 r_0^2} \biggr) \, . \label{Q0coeff}
\end{equation}
For $q_0 = 0$ \emph{and} $u_0^{(1)} = 0$, these expressions agree
exactly with Eqs. (17) and (18) in Ref. \cite{AMS2}. However, the
zeroth--harmonic correction $u_0^{(1)}$ \emph{cannot} be set equal
to zero; indeed, collecting the constant (zeroth--harmonic) terms
arising in this order, one obtains
\begin{equation}
\frac{\partial^2 u_0^{(1)}}{\partial \zeta^2} \, = \,- \frac{p_0
k^2}{v_g^2 - v_0^2} \, \frac{\partial }{\partial \zeta} \,
|u_1^{(1)}|^2 \, , \label{NLSE02a}
\end{equation}
which may immediately be integrated, yielding
\begin{equation}
\frac{\partial u_0^{(1)}}{\partial \zeta} \, = \,- \frac{p_0
k^2}{v_g^2 - v_0^2} \, |u_1^{(1)}|^2 \, + \, C \, \equiv R(k)\,
|u_1^{(1)}|^2 \, + \, C \, , \label{NLSE02}
\end{equation}
where $C$ is a constant, to be determined by the boundary conditions
(e.g. $C = 0$ if $u_1^{(1)} = 0$ at
$x = \pm \infty$). Note that the
resonance condition $v_g = v_0$ amounts to $k^2 - 3 v_0^2 /(4
v_1^2 r_0^2) = 0$, which is \emph{not} satisfied in the parameter
range of the continuum approximation; on the contrary, $v_g < v_0$
everywhere, implying that $R(k)$ takes positive/negative values
for positive/negative $p_0$: for Debye interactions, $p_0 > 0$ and
$R > 0$. The non--vanishing constant term $u_0^{(1)}$ is purely
due to cubic interaction potential nonlinearity (and disappears
for $p_0 = 0$). This is known in
solid state physics \cite{Tsurui, Flytzanis1985}.

Eqs. (\ref{NLSE01}) and (\ref{NLSE02}) form a coupled set of
evolution equations, governing the dynamics at first order in
$\epsilon$. They may readily be combined into the form of a
Nonlinear Schr\"odinger Equation (NLSE)
\begin{equation}
i \frac{\partial A}{\partial \tau} \, + \, P \, \frac{\partial^2
A}{\partial \zeta^2} \, + \,Q \, |A|^2 A\,  = 0\, ,  \label{NLSE}
\end{equation}
which, once solved for the amplitude of the first harmonic $A =
u_1^{(1)}$, immediately provides the solution for $u_0^{(1)}$ upon
substitution into (\ref{NLSE02}). A linear term $\sim C \,A$ has been
neglected, since it is canceled by a trivial phase shift
transformation. The final form of the nonlinearity coefficient $Q$
now reads
\begin{equation}
Q \, = \,- \frac{k^2}{2 \omega} \, \biggl(q_0 \,k^2 +
\frac{p_0^2}{6 v_1^2 r_0^2}\biggr)  \, - \, \frac{p_0^2 k^4}{2
\omega} \, \frac{1}{v_g^2 - v_0^2} \, . \label{Qcoeff}
\end{equation}

%The coefficients $P$ and $Q$ are depicted in Fig. \ref{figurex}.
In the infinite wavelength limit (i.e. near $k \approx 0$), we
have
\[P \approx - \frac{3 v_1^2 r_0^2}{2 v_0} k - \frac{5 v_1^4 r_0^4}{4 v_0^3} k^3
+ {\cal O}(k^5)
\]
and
\[Q \approx  \frac{p_0^2}{12 v_1^2 r_0^2 v_0} k + \biggl(
\frac{7 p_0^2}{72 v_0^3} - \frac{q_0}{2 v_0} \biggr) k^3 + {\cal
O}(k^5) \, .
\]
Let us stress that the extra term, which makes the difference
between $Q_0$ and $Q$, changes completely the behaviour of the
nonlinearity coefficient near $k \approx 0$. For example, we find
that
\[Q_0 \approx - \frac{p_0^2}{12 v_1^2 r_0^2 v_0} k - \biggl(
 \frac{p_0^2}{24 v_0^3} + \frac{q_0}{2 v_0} \biggr) k^3 + {\cal
O}(k^5) \, .
\]

Note, once more, that the results in this Section are general and
apply to a wide class of physical problems. In any given problem,
involving an arbitrary (analytical, possibly long--range)
interaction law, Eqs. (\ref{NLSE01}), (\ref{NLSE02}) and
(\ref{NLSE}) provide the evolution of the first--order correction
$u_1$ to the particle displacement $u$, if the latter is governed
by an equation of motion of the form of (\ref{eqmotion1}), as
expressed in the continuum limit by Eq.
(\ref{eqmotion-gen-continuum}). To first order in $\epsilon$, the
solution then reads
\[
u(x, t) \approx \, \epsilon \,[u_0^{(1)}(\zeta, \tau) + \,
u_1^{(1)}(\zeta, \tau) \,+ c.c.] + \mathcal{O}(\epsilon^2) \, ,
\]
where $u_0^{(1)}$ is given by (\ref{NLSE02}) (it vanishes if, and
only if, $p_0 = 0$) and $u_1^{(1)}$ corresponds to (the amplitude
of) a small translation and an oscillation, respectively, both
traveling at the group velocity $v_g$. We emphasize, as a general
result, that the existence of a non--zero cubic potential
nonlinearity term $p_0$ implies the existence of a finite
translation (zeroth--harmonic) correction $u_0^{(1)}$ to first
order in $\epsilon$, while the sign of $p_0$ affects the form of
$u_0^{(1)}$ (i.e. the lattice compression or rarefaction);
furthermore, it also determines the (sign of the) second--harmonic
correction, as we saw above. Nevertheless, the sign of $p_0$ does
not affect the form of the coefficients in (\ref{NLSE}), and thus
neither the plane wave stability profile.

\section{Stability analysis}

The standard stability analysis \cite{Remoissenet, Hasegawa, Newell}
consists in linearizing around the (Stokes) plane wave solution
of (\ref{NLSE}), \(A \, = \, {\hat A} \, e^{i Q |\hat A|^2
\tau}  +  c.c.  \) (notice the amplitude dependence of the
frequency), by setting \({\hat A}  =  {\hat A}_0 + \epsilon \,
{\hat A}_1 \) and taking the perturbation ${\hat A}_1$ to be of
the form ${\hat A}_1 \, = \, {\hat A}_{1, 0} \,e^{i ({\hat k}
\zeta - {\hat \omega} \tau)} \, + \, c.c.$, where $\hat k$ and
$\hat \omega$ denote the perturbation's wavenumber and frequency,
to be distinguished from the carrier wave's homologous quantities,
$k$ and $\omega$. Hence from (\ref{NLSE}) one readily
obtains the dispersion relation
%\begin{equation}
\( \hat \omega^2 \, = \, P^2 \, \hat k^2 \, \biggl(\hat k^2 \, -
\, 2 \frac{Q}{P} |\hat A_{0}|^2 \biggr) \).
%\, .\end{equation}
The wave will thus be {\em stable}
%($\forall \, \hat k$)
if the
product $P Q$ is negative. However, for positive $P  Q > 0$,
instability sets in for wavenumbers below a critical value $\hat
k_{cr} = \sqrt{2 \frac{Q}{P}} |\hat A_{0}|$, i.e. for wavelengths
above a threshold: $\lambda_{cr} = 2 \pi/\hat k_{cr}$, with an increment
\( \sigma = |Im\hat\omega(\hat k)| \)
which attains a maximum value at $\hat k = \hat
k_{cr}/\sqrt{2}$, viz.
\[ \sigma_{max} =
|Im\hat\omega|_{\hat k = \hat k_{cr}/\sqrt{2}} \,=\, | Q |\, |\hat
A_{0}|^2  \, .
%\label{growthrate}
\]
It turns out that the instability condition depends only on
the sign of the product $P Q$, which may be studied numerically
by using the expressions derived above.

Let us go back to the expressions (\ref{Pcoeff}) and (\ref{Qcoeff}) for
the coefficients $P$ and $Q$. First, we notice that $P$ \emph{is
always negative}, given the condition $k < k_{cr, 1} = (v_0/v_1)
\, r_0^{-1}$ imposed above. On the other hand, the nonlinearity
coefficient $Q$ may take either negative or positive values,
depending on the physical parameters $v_0$, $v_1$, $p_0$ and $q_0$
and, of course, the wavenumber $k$. To see this, one may cast
(\ref{Qcoeff}) in the form
\begin{equation}
Q \, = \, - \frac{k^2}{2 \omega}\, \biggl[ q_0 \, k^2 \, - \,
\frac{3 - 2 v_1^2 r_0^2 k^2/v_0^2}{3 - 4 v_1^2 r_0^2 k^2/v_0^2} \,
\frac{p_0^2}{6 v_1^2 r_0^2} \biggr] \label{Qcoeff-reduced} \, ,
\end{equation}
and then examine its sign in terms of $k$. This can be done either
by a straightforward (yet rather lengthy) analytical calculation
(omitted here) or numerically (see below).
Notice, as a general remark, the pole created at the resonance
point $v_g = v_0$, for $k = v_0 \sqrt{3}/(2 v_1) \, r_0^{-1}$
($= 3 \, r_0^{-1}$ for FNI).
In conclusion, the
carrier wave will be modulationally \emph{unstable} for $k$ values
corresponding to $Q < 0$ ($P Q > 0$), and \emph{stable} for $Q >
0$ ($P Q < 0$). These results are valid for any given form of the
electrostatic interaction potential $\phi(r)$ and neighboring site
vicinity $N$, which (both) in fact enter the physical parameters
via the expressions (\ref{defv0}) -- (\ref{defc111}).

\begin{figure}[htb]
 \centering
  \resizebox{4in}{!}{
 \includegraphics[]{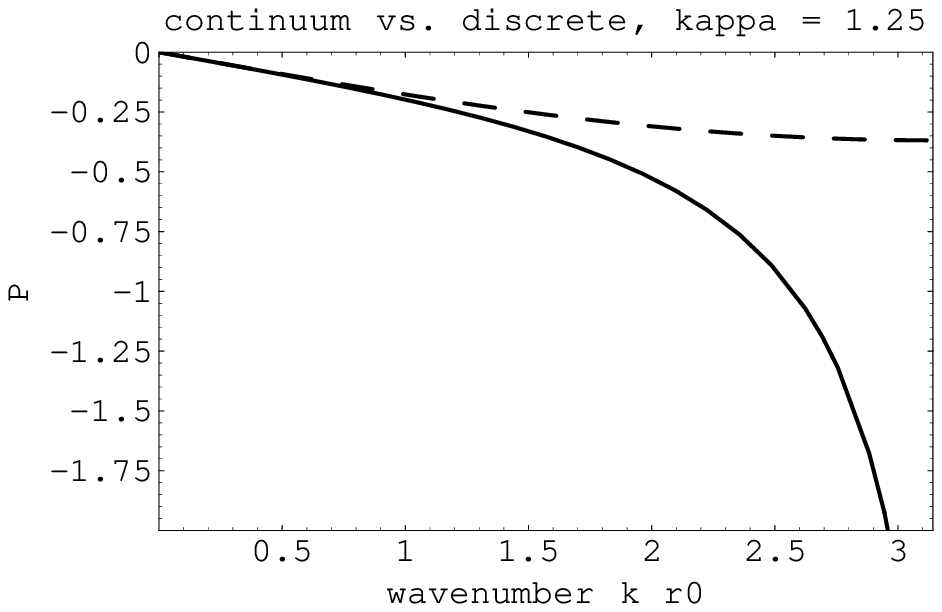}} \\
\vskip 2 cm \resizebox{4in}{!}{
 \includegraphics[]{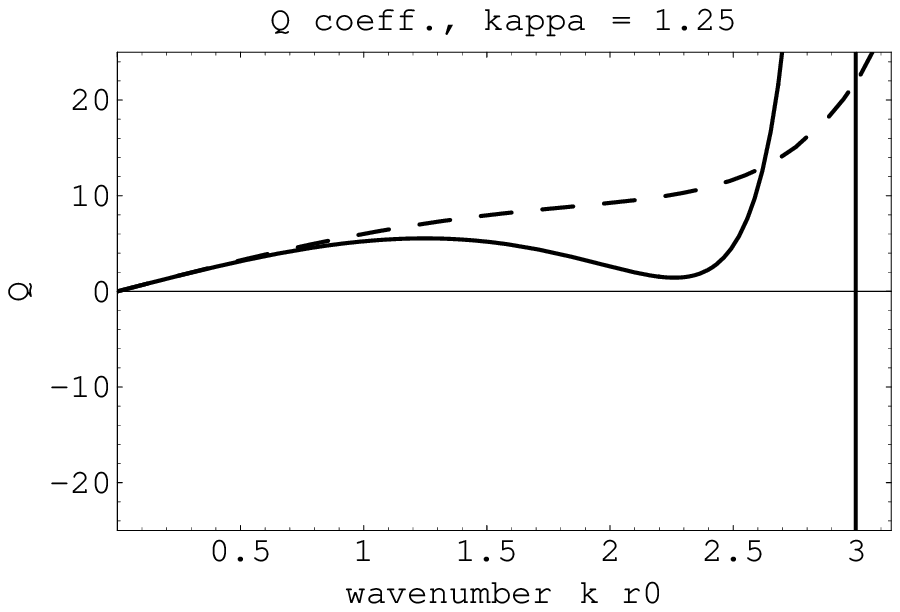}}\\
 \vskip 2 cm
 \resizebox{4in}{!}{
 \includegraphics[]{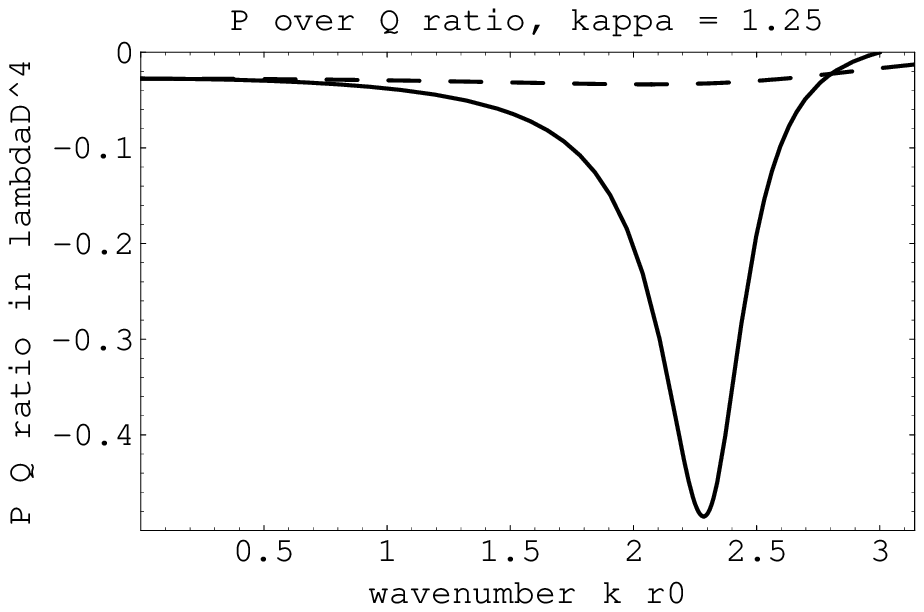}}
\caption{} \label{figure3}
\end{figure}

\begin{figure}[htb]
 \centering
  \resizebox{4in}{!}{
 \includegraphics[]{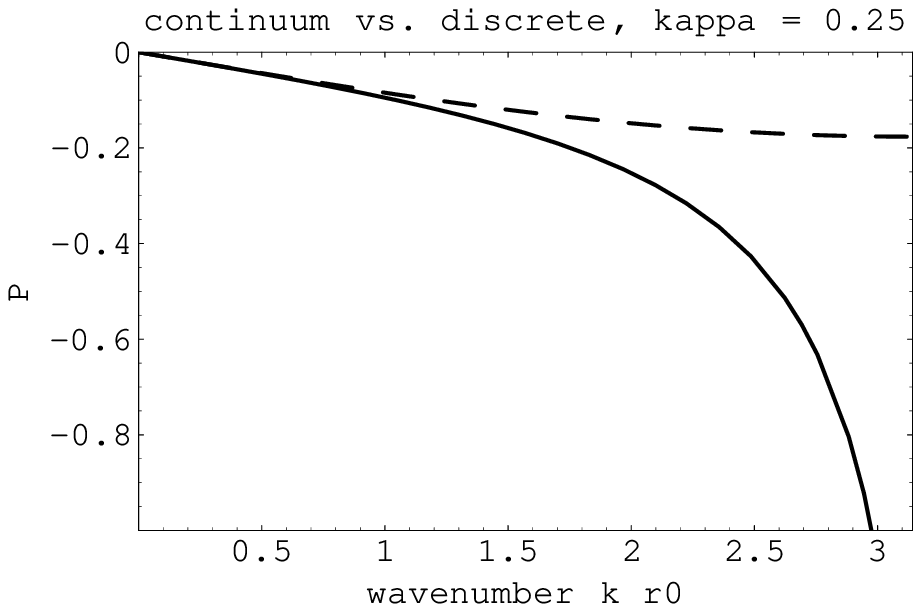}} \\
\vskip 2 cm \resizebox{4in}{!}{
 \includegraphics[]{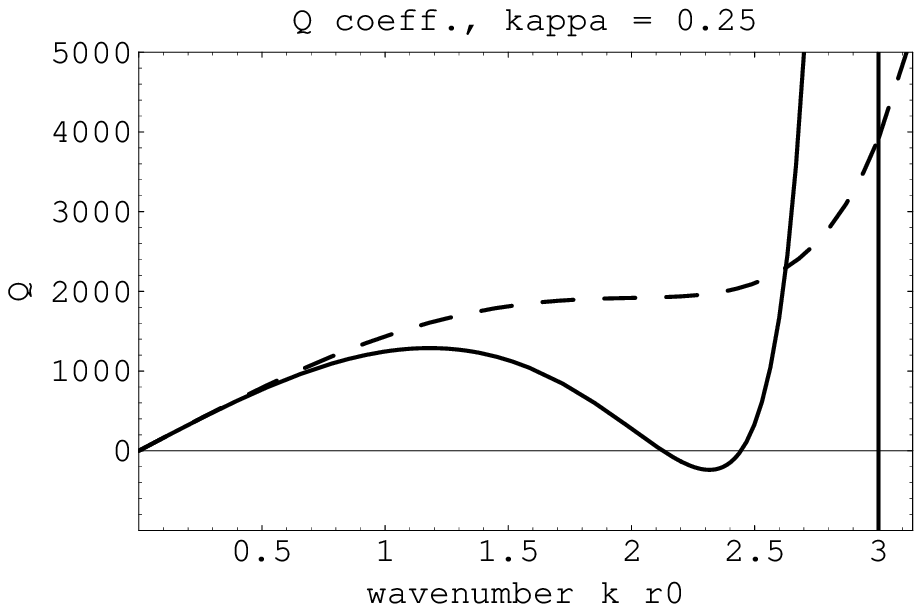}}\\
 \vskip 2 cm
 \resizebox{4in}{!}{
 \includegraphics[]{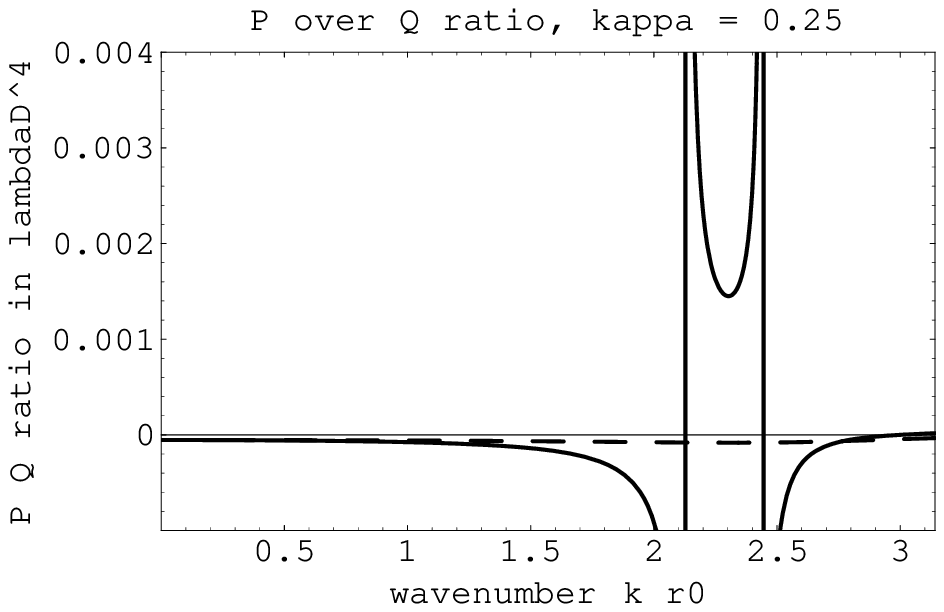}}
\caption{} \label{figure4}
\end{figure}

Let us now focus on the dust--lattice problem. At a first step,
one may explicitly assume first neighbor Debye--H\"uckel type
interactions, by substituting with expressions (\ref{Debye-om0})
-- (\ref{Debye-q0}) into the above relations for $P$, $Q$, and
then study the behaviour of all the relevant physical quantities as a
function of, say, the lattice parameter $\kappa$. Notice that the
stability profile deduced above is dramatically modified, in
comparison with the limit of vanishing cubic interaction nonlinearity,
i.e. $q_0 = 0$, considered in Ref. \cite{AMS2}, where the
instability of plane waves was directly prescribed (in
addition to $u_0^{(1)} = 0$). A simple numerical investigation shows
that $Q$ is always {\em{positive}} (prescribing stability, since $P < 0$)
at low values of the wavenumber
$k$; see
Figures \ref{figure3}, \ref{figure4}.
This is true everywhere, for $k$ below the resonance threshold
$k r_0 \approx 3$, except for $\kappa$ of the order of
$\approx 0.75 \,r_0^{-1}$
or lower, where a small instability region is encountered; see
Fig. \ref{figure4} [notice the asymptotic curve
at $k r_0 = v_0 \sqrt{3}/(2 v_1) = 3$, as expected from
(\ref{Qcoeff-reduced}), beyond which the continuum approximation
certainly fails]. Long wavelength LDLWs are therefore
expected to be stable, while shorter wavelength LDLWs may be
unstable. For rigor, it should be admitted that the instability
(carrier wavenumber) threshold $k_{cr}$ is quite high, and rather
beyond the validity of this continuum model
(recall the assumption made above, that $k \ll k_{cr, 0} = \pi/r0$).
Therefore, while small $k$ stability may be taken for
granted, we only possess an indication for (i.e. not a rigorous
proof of) instability for longer $k$. We draw the conclusion that
LDL waves are most likely to remain stable with respect to
external perturbations (even though this statement may a priori
not be valid for a different interaction potential).
The present results are in agreement with previous prediction in atomic chains
\cite{Tsurui, Flytzanis1985}.

\section{Nonlinear excitations}

Equation (\ref{NLSE}) is an integrable nonlinear partial
differential equation \cite{Newell, ZS}, which is known to possess
different types of localized, constant profile, travelling wave
solutions (solitons). Several types of such localized modes are
presented in Refs. \cite{Hasegawa, ZS, Flytzanis1985, Fedele})
(see Ref. \cite{Fedele} for a review) so we will only briefly
outline the analytical form of those we are interested in, and
discuss their relevance to our problem.

A solution of Eq. (\ref{NLSE}) may be sought in the form
\(A(\zeta, \tau) = \rho(\zeta, \tau) \, e^{i\,\Theta(\zeta, \tau)
} \), where $\rho$ and $\sigma$ are real variables to be
determined. Different types of solution are thus obtained,
depending on the sign of the product $P Q$. Once \(u_1^{(1)} \, =
A\) is thus determined, the zeroth--harmonic first--order
correction $u_0^{(1)}$ is obtained from (\ref{NLSE02}).

%\subsection{Asymmetric bright solitons}

\begin{figure}[htb]
 \centering
 \resizebox{4in}{!}{
 \includegraphics[]{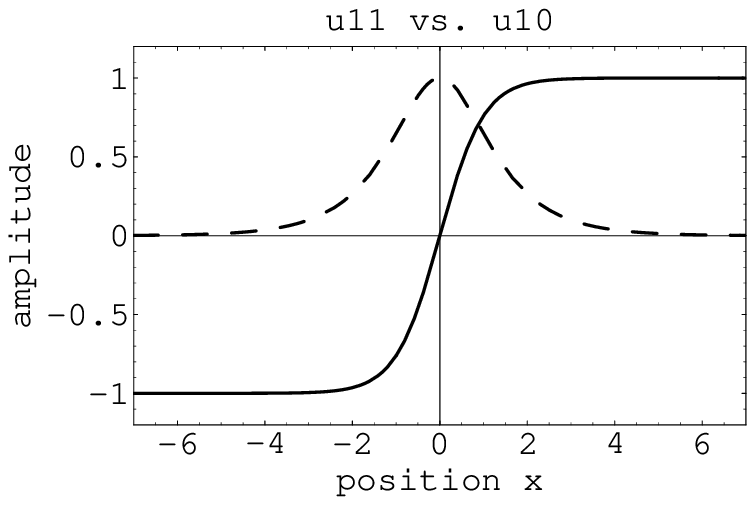}}\\
\vskip 2 cm \resizebox{4in}{!}{
 \includegraphics[]{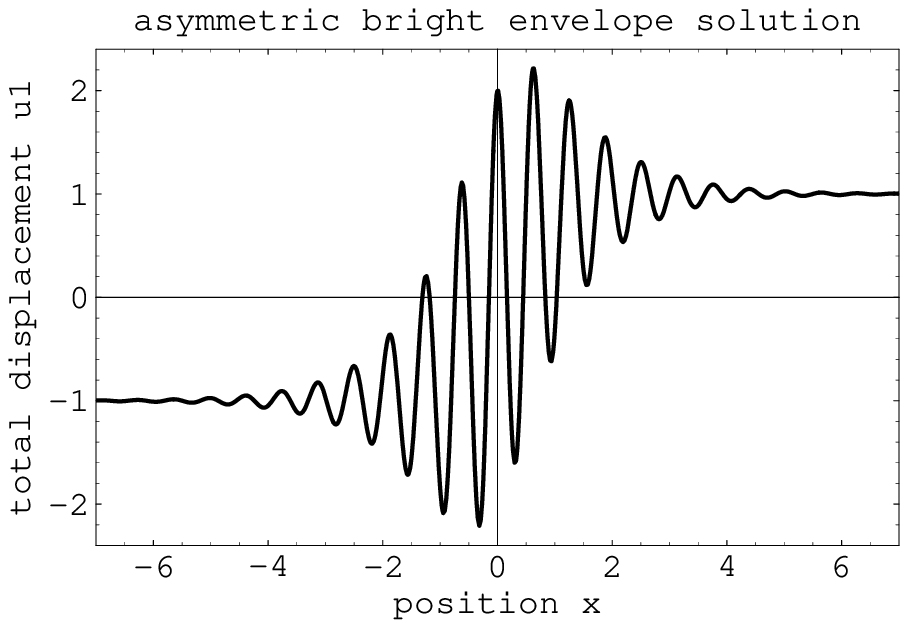}}
\caption{} \label{figure5}
\end{figure}

For $P Q > 0$ we find the {\em bright envelope soliton} \cite{commentFedele1}
\begin{equation}
A(x, t)\, = \, \rho_0 \, sech[(x - v_e t)/L_e + x_e]\, \exp [ - 2
\, i\, \alpha\, (x - v_g t) \, -\, 4 (\alpha^2 - \eta^2) \, P \, t
]  \, \label{bright}
\end{equation}
($sech x = 1/\cosh x$), which represents a localized envelope
pulse travelling at a speed $v_e = v_g - 4 \alpha \,P$. The pulse
width $L_e = 1/(2 \eta)$ is related to the (constant) maximum
amplitude $\rho_0 = 2 \,\eta\, (P/Q)^{1/2}$ as $\rho_0 =
({P}/{Q})^{1/2}/L_e$. See that both of the small (arbitrary)
parameters $\alpha$ and $\eta$ bear dimensions of inverse length.
Substituting (\ref{bright}) into (\ref{NLSE02}), we obtain the zeroth--harmonic
term $u_0^{(1)} = R\, \int |u_1^{(1)}|^2 \, d\zeta =\, 2 \, \eta\,
P R(k)/Q \, \tanh[(x - v_e t)/L_e + x_e] + c$
(we will assume that
$c = 0$). Combining the preceding formulae, we obtain the
\emph{asymmetric} bright--envelope modulated solution for the
(total) dust--grain displacement (to order $\sim \epsilon$)
\begin{equation}
u_1 \, = \, A_0 \, sech[(x - v_e t)/L_e + c_e]\, \cos[(x-v_c t
)/L_0 + c_0] \,+ \,B_0 \,\tanh[(x - v_e t)/L_e +
c_e] \label{bright-asymmetric}
\end{equation}
where $A_0 = 2 \rho_0 = 4 \eta (P/Q)^{1/2}$, $B_0 = {R(k)}
\,\rho_0^2\, L_e = 2 \, \eta\, R(k) \, P / Q$ and the shifted
(modulated) carrier wavenumber $k_c = L_0^{-1}$ and frequency
$\omega_c = v_0/L_0$ now depend on: $L_0 = 1/(k - 2 \alpha)$ and
$v_c = [\omega - 2 \alpha v_g + 4 P\, (\alpha^2 - \eta^2)] \,
L_0$; $c_e$ and $c_e$ are arbitrary real constants. Equation
(\ref{bright-asymmetric}) represents a superposition of a
localized modulated harmonic oscillation and a kink--like
excitation, which is characterized by different constant
asymptotic limits (at $x = \pm \infty$). Since $R(k) > 0$ in a
Debye crystal, this solution will therefore represent a localized
compression co--propagating with a strongly modulated envelope
oscillation; see Fig. \ref{figure5}.

%\subsection{Asymmetric dark solitons}

\begin{figure}[htb]
 \centering
 \resizebox{4in}{!}{
 \includegraphics[]{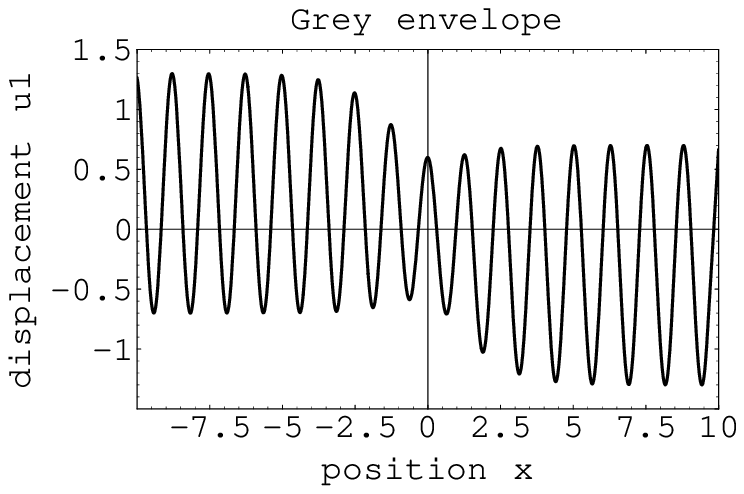}}\\
\vskip 2 cm \resizebox{4in}{!}{
 \includegraphics[]{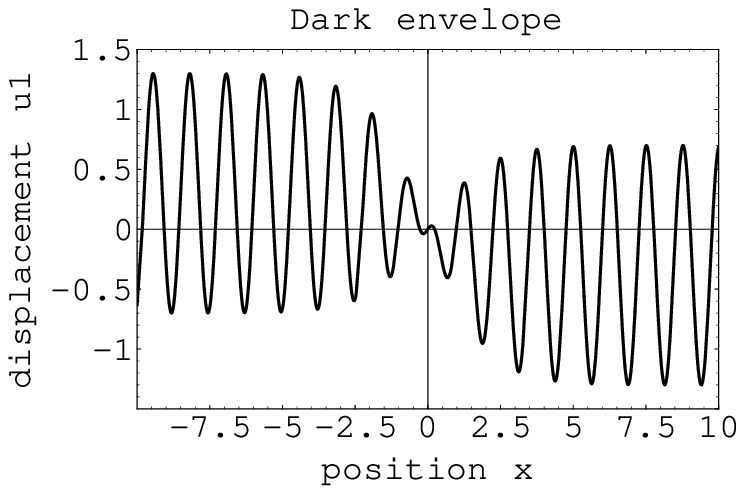}}
\caption{} \label{figure6}
\end{figure}

For $P Q < 0$, we have the {\em grey} envelope soliton \cite{commentFedele1}
\begin{equation}
A(\zeta, \tau) \, = \, \rho_1 \, \{ 1 - a^2 \, sech^2[(\zeta - 2 \alpha
\tau)/L_1]\}^{1/2} \, \exp[ i\, \sigma(\zeta, \tau)] \, ,
\label{greysoliton}
\end{equation}
where \[\sigma(\zeta, \tau) = \sin^{-1} \{ a \tanh (\zeta/L_1)\, [
1 - a^2 \, sech^2(\zeta/L_1) ]^{-1/2} \, - 2 \alpha \zeta +
(\Omega - 4 P \alpha^2)\tau
 \}
\]
which represents a localized region of negative wave density (a
void), with finite amplitude $(1 - a) \rho_1$ at $\zeta = 0$; $0
\le a \le 1$) [one may choose $\Omega = 4 P \eta^2$ and $\rho_1 =2
\, \eta\, (|P/Q|)^{1/2}$, for analogy with (\ref{bright})]. Again,
the pulse width $L_1 = (|P/Q|)^{1/2}/(a \rho_1)$ is inversely
proportional to the amplitude $\rho_1$. Going back to the original
variables, we obtain the dust grain harmonic oscillation
\begin{equation}
u_1^{(1)}\,\exp i\theta + c.c.\, = \, A_1 \, \{ 1 - a^2 \,
sech^2[(x - v_e t)/L_1
+ c_e]\}^{1/2} \, \cos[(x - v_c t)/L_0 + \, \sigma_1(x, t)] \, ,
\label{greyu11}
\end{equation}
which has a maximum amplitude $A_1 = 2 \, \rho_1$ and a phase shift
 \[ \sigma_1(x, t) = \sin^{-1} [ a \tanh (x - v_e t)/L_1]\, [
1 - a^2 \, sech^2(x - v_e t)/L_1 ]^{-1/2} \, + \sigma_0 \, ,
\]
where $v_e$, $v_c$ and $L_0$ are just as defined previously.
Notice the (dimensionless) parameter $a$, which regulates the
depth of the excitation. For $a = 1$, one obtains the {\em dark}
envelope soliton
\begin{equation}
u_1^{(1)}\exp i \theta + c.c. \, = \, A_1 \, \tanh[(x - v_e t)/L_1
+ c_e] \, \cos[(x - v_c t)/L_0 + \, c_0] \,,
%\label{greysoliton}
\end{equation}
which describes a localized density \textit{hole}, characterized
by a vanishing amplitude at $\zeta = 0$; see Fig. \ref{figure6}.
The dark-- or grey--soliton--modulated wavepackets in the previous
paragraph, are superposed on a constant (zeroth--harmonic)
displacement, given by (\ref{NLSE02}) as
\begin{equation}
u_0^{(1)}\, = \, - B_1\, \tanh[(x - v_e t)/L_1
+ c_e] \, ,
\label{greyu01}
\end{equation}
where $B_1 = R(k)\, \rho_1^2 \, a^2 \, L_1 \, = \, a\, R(k)\,
(|P/Q|)^{1/2}\, \rho_1$; the constant $C$ in (\ref{NLSE02}) was
chosen as $C = - R(k) \, \rho_1^2$ for constant asymptotic values
at infinity. Combining the latter formulae, we obtain the
\emph{asymmetric} dark/grey modulated envelope solution for the
dust--grain total displacement (to order $\sim \epsilon$)
\begin{equation} u_1 \, = \, A_1 \, \{ 1 - a^2 \, sech^2[(\zeta - 2 \alpha
\tau)/L_1]\}^{1/2} \, \cos[(x-v_c t )/L_0 + \sigma + c_0] \,-
\,B_1 \,\tanh[(x - v_e t)/L_e + c_e] \label{grey-asymmetric}
\end{equation}
or, for $a = 1$,
\begin{equation} u_1 \, = \, A_1 \, \tanh[(x - v_e t)/L_1 +
c_e]\, \cos[(x-v_c t )/L_0 + c_0] \,- \,B_1 \,\tanh[(x - v_e
t)/L_e + c_e] \label{dark-asymmetric}
\end{equation}
where all quantities were defined previously. For $R > 0$, as in
the Debye DP crystal case, the dark/grey type excitations
correspond to a localized density dip accompanied by a rarefaction
propagating in the lattice; see Fig. \ref{figure6}.

\begin{figure}[htb]
 \centering
 \resizebox{4in}{!}{
 \includegraphics[]{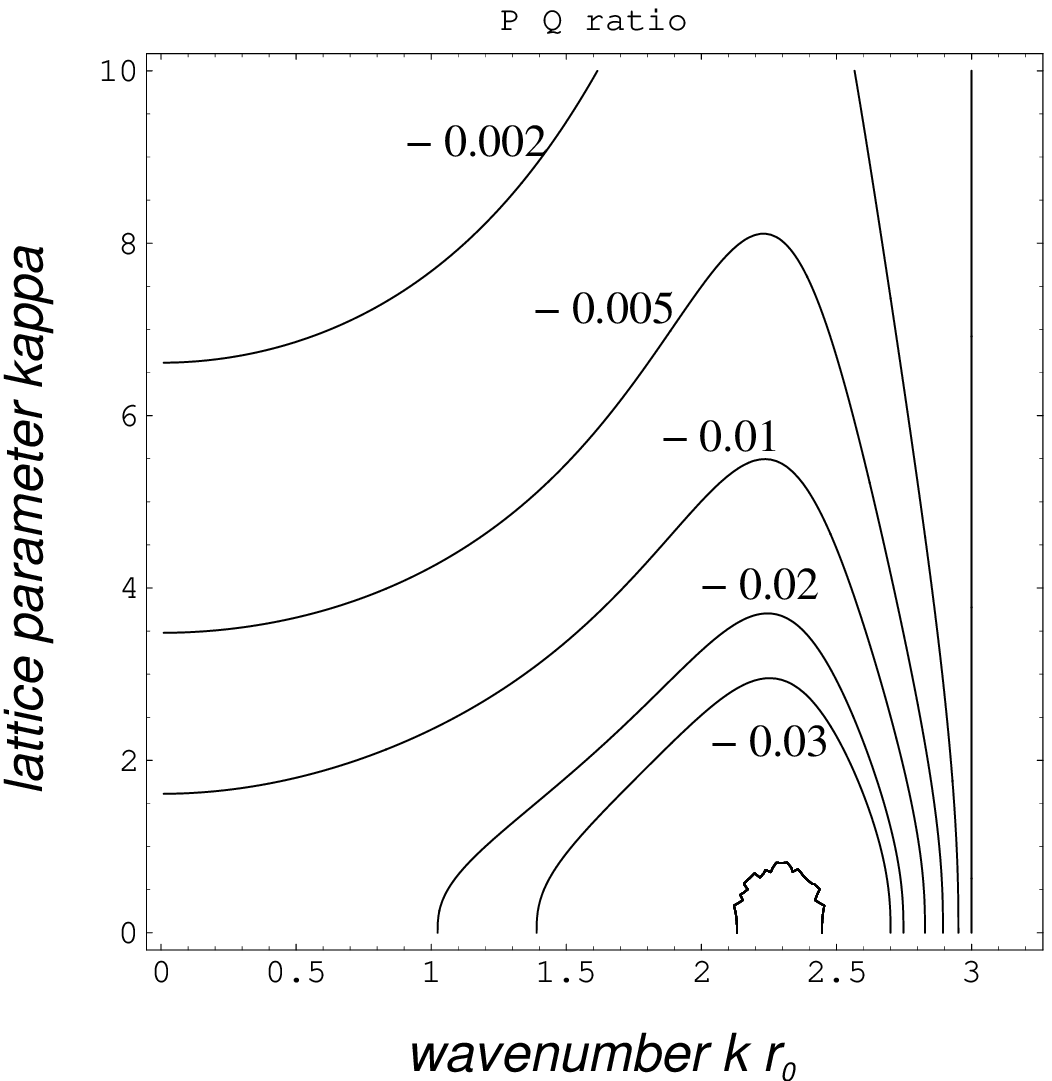}}\\
\vskip 2 cm \resizebox{4in}{!}{
 \includegraphics[]{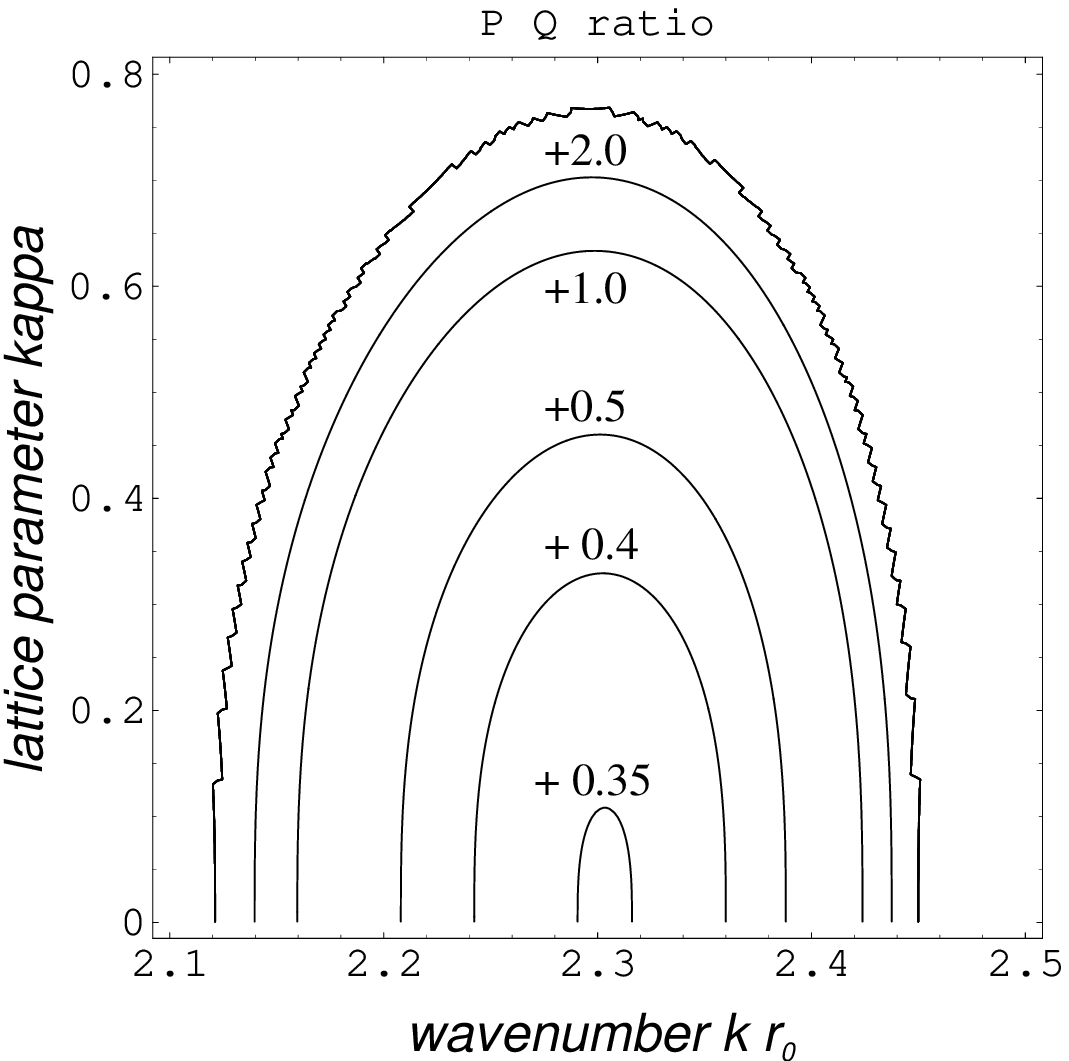}}
\caption{} \label{figure7}
\end{figure}

A quantity of importance comes out to be the ratio $\xi = P/Q$:
its sign dictates the type of envelope soliton which should, in
principle, propagate in the lattice, while (the square root of)
its absolute value represents a measure of the soliton amplitude
$A_0$ for a given width $L$, viz. $A_0 = \sqrt{\xi}/L$. In figure
\ref{figure7}, we have depicted $\xi$ versus the (normalized)
wavenumber $k r_0$ and the lattice parameter $\kappa$. We notice
an increase of the absolute value of the (negative, mostly) ratio
with decreasing $\kappa$ and with increasing $k$. One draws the
conclusion that for a given excitation width $L$ (assumed $L \gg
r_0$, according to our approximation), shorter wavelength (i.e.
higher wavenumber) carrier waves will form dark--type excitations
(pulses) with an increased maximum displacement $A_0$; this
picture, which is true below, say, $k \approx 2.25 \, r_0^{-1}$,
is inverted above this value (nevertheless, there may be some
doubt above the validity of results in that region, as discussed
above). Also, higher values of $\kappa$, i.e. higher lattice
constant $r_0$ values for a given Debye length $\lambda_D$, seem
to favor narrower excitations (as compared to $r_0$). Notice, in
Fig. \ref{figure7}a, the curly--shaped curve at the bottom,
indicating the poles of the ratio $P/Q$, i.e. where $Q = 0$ ($\xi
> 0$ only below this curve), as well as the resonance asymptotic
line at $k r_0 = 3$. See that positive values of $\xi$ are
confined in a small, rather short--wavelength and small $\kappa$
region; see Fig. \ref{figure7}a. As a final comment, notice that
values deduced from Fig. \ref{figure7} are in agreement with the
continuum approximation. For instance, as a crude estimation,
taking $L = 5 r_0$ and $\kappa \approx 1$, one obtains $\xi
\approx 0.025 r_0^2$, implying $A_0 = \sqrt{\xi}/L \approx 0.03
r_0 $, i.e. $A_0 \ll r_0$ as predicted.

Concluding, the instability/stability regions depicted e.g. in
Figs. \ref{figure3}, \ref{figure4}, in fact also limit the ($k,
\kappa$) parameter pair values where bright (dark/grey) solutions
i.e. density pulses (holes) may exist. Furthermore, the envelope
characteristics will depend on the carrier wave dispersion via $P$
and $Q$; for instance, regions with lower values of $P$ (or higher
values of $Q$) will support narrower excitations. Admittedly,
however, dust lattices will rather not favor bright--type
structure formation by unstable long--wavelength LDL waves. On the
other hand, such waves will mostly be stable to external
perturbations, and may locally form density dips (envelope holes).

\section{Conclusions}

In this paper, we have investigated the amplitude modulation of
longitudinal dust-lattice waves propagating in a one-dimensional
dust-crystal. By using a continuum approximation, we have derived
a Nonlinear Schr\"odinger Equation governing the evolution of the
LDL wave envelope. We have found that long wavelength carrier
waves will generally be stable, possibly evolving towards the
formation of voids, i.e. asymmetric dark/grey--type coherent
envelope structures, while longer wavelengths may allow for the
formation of envelope pulses (asymmetric bright--type envelope
excitations). Furthermore, we observe that the total grain displacement will
bear a (weak) constant displacement (zeroth mode), due to the
intrinsic third--order nonlinearity of the Debye potential. It may
be admitted that the former (dark) excitations may rather not be
relevant to this idealized, infinite chain model, since they
correspond to an infinite amount of energy stored in the lattice.
Nevertheless, they may be taken as a hint towards excitations
sustained in finite-length, highly discrete, real lattices.

The present study aims in making a first step towards the
elucidation of the nonlinear harmonic generation mechanisms
dominating weakly--nonlinear longitudinal wave packets in
dust-crystals. The formation of LDLW--related structures has thus
been predicted, and will hopefully be confirmed by appropriately
designed experiments or by numerical simulations. In fact, through
the present investigation, a link has been established towards the
formalism of discreteness--related intrinsic localized modes
(discrete breathers) widely studied recently in solid state
systems \cite{breathers, Kiselev}. This work should, in principle,
be extended by including realistic effects associated with crystal
asymmetries, defects, dust charging, ion-drag and multiple
dust-layer coupling \cite{Ivlev}. In particular, one should
consider a similar effect in more realistic, e.g. 2D, geometries,
as the ones observed in recent experiments with laser-excited dust
lattices \cite{Nunomura, Samsonov}. Concluding, this work was
devoted to a nonlinear mechanism for harmonic generation and
coherent structure formation. Although here presented in a 1D
geometry, for simplicity, this qualitative mechanism may be
generalized to 2D lattices; this possibility will be considered in
future work.

\begin{acknowledgments}
This work was partially supported by the European Commission
(Brussels) through the Human Potential Research and Training
Network via the project entitled ``Complex Plasmas: The Science of
Laboratory Colloidal Plasmas and Mesospheric Charged Aerosols''
(Contract No. HPRN-CT-2000-00140).
\end{acknowledgments}

\newpage

% figure CAPTIONS

\centerline{\textbf{Figure captions}}

Figure 1.

(a) The characteristic physical quantities defined in (\ref{Debye-om0}) --
 (\ref{Debye-q0}) (for Debye--type first neighbour interactions)
are depicted against the lattice parameter $\kappa$: the
eigenfrequency (square) $\omega_{L, 0}^2$, normalized by $Q^2/M
\lambda_D^3$ (solid curve ---); the sound speed (square)
$v_{0}^2$, normalized by $Q^2/M \lambda_D$ (long-dashed curve --
-- --); the cubic potential nonlinearity coefficient $p_0$,
normalized by $Q^2/M \lambda_D$ (dot curve $\cdot \cdot \cdot$);
the quartic potential nonlinearity coefficient $q_0$, normalized
by $Q^2/(M \lambda_D)^{1/2}$; short-dashed curve -  - -); (b)
close-up near $\kappa \approx 1$.

\medskip

Figure 2.

(a) The linear oscillation frequency $\omega$ (normalized byr
$\hat \omega_0 = (2 Q^2/M \lambda_D^3)^{1/2}$) is depicted against
the normalized wavenumber $k r_0$; the three curves, from bottom
to top, depict: the exact relation (\ref{dispersion-discrete})
(---), the continuum approximation (\ref{dispersion}) (-- -- --),
and the (tangent) acoustic limit $\omega = k v_0$  (- - -); (b)
close-up near the origin.

\medskip

Figure 3.

The NLS coefficients $P$, $Q$ and their ratio $\xi = P/Q$:
discrete model [see (\ref{dispersion-discrete})] (dashed curve)
vs. continuum model (straight curve), are depicted vs. the
(normalized) wavenumber $k r_0$, for $\kappa = 1.25$. (a) $P$, cf.
(\ref{Pcoeff}) (normalized by $(Q^2 \lambda_D/M)^{1/2}$); (b) $Q$,
cf. (\ref{Qcoeff}) (normalized by $[Q^2/(M \lambda_D^7)]^{1/2}$);
(c) $\xi = P/Q$ (normalized by $\lambda_D^{4}$; divide by a factor
$\kappa^4 = 1.25^4 \approx 2.44$ to obtain the value in $r_0^4$).

\medskip

Figure 4.

Similar to the previous figure, for $\kappa = 0.25$. The continuum
model seems to fail completely above $k = 1$: notice the discrete
(correct) curve, which is hardly visible (dashed curve).

\medskip

Figure 5.

(a) Bright (asymmetric) envelope solution ($P Q > 0$): the first--
and zeroth-- harmonic components of the particle displacement,
$u_1^{(1)}$ and $u_0^{(1)}$, are depicted, for $t=0$; dash (-- --
-- ) and solid (---) curves, respectively. (b) The total
displacement $u_1$ ($\sim \epsilon$) is depicted (arbitrary
parameter values).

\medskip

Figure 6.

Asymmetric envelope solutions for $P Q < 0$: the total
displacement is depicted for $t=0$ (arbitrary parameter values):
(a) grey--type envelope solution ($0 < |a| < 1$); (a) dark--type
envelope solution ($|a| = 1$).

\medskip

Figure 7.

Contours of the reduced ratio $\xi = P/Q$ (expressed in units
$r_0^4$) are depicted vs. the (normalized) wavenumber $k r_0$ and
the lattice parameter $\kappa$. (a) The $\xi$ values are, from top
left and downwards: $-0.002$, $-0.005$, $-0.01$, $-0.02$, $-0.03$.
The absolute value of the (negative) ratio increases with
decreasing $\kappa$ and with increasing $k$.
Notice the curly--shaped curve at the bottom, indicating
the poles of the ratio $P/Q$, i.e. where $Q = 0$ ($\xi > 0$ below
this curve), as well as
the resonance asymptotic line at $k r_0 = 3$.
(b) Close up for small $\kappa$ and positive $\xi$
(wave instability region).
The $\xi$ values are, from bottom to
top: $0.35$, $0.4$, $0.5$, $1$, $2$ (and $+\infty$, again, i.e. $Q=0$, on
top).

\newpage

\vskip 2 cm

% figure1

\newpage

\vskip 2 cm

% figure2

\newpage

\vskip 2 cm

% figure3

\newpage

\vskip 2 cm

% figure4

\newpage

\vskip 2 cm

% figure5

\newpage

\vskip 2 cm

% figure5

\newpage

\vskip 2 cm

% figure7


\begin{thebibliography}{10}
\bibitem{commentFedele1} These expressions may be obtained
from the ones in Ref.
\cite{Fedele}, by transforming the variables therein into our
notation as: $\rho \rightarrow \rho^2$,
$x \rightarrow \zeta$, \, $s \rightarrow
\tau$, \, $\rho_m \rightarrow \rho_0^2$, \ $\alpha \rightarrow 2 P$,
\ $q_0 \rightarrow - 2 P Q$, \ $\Delta \rightarrow L$, \ $E
\rightarrow \Omega$, \ $V_0 \rightarrow V$; also see Ref. \cite{Flytzanis1985}.
\end{thebibliography}
\end{document}